\documentclass[10pt]{article}

\usepackage{siunitx}
\usepackage{amsmath}
\usepackage{natbib}
\usepackage{multirow}
\usepackage{graphicx}

\title{Bayesian Stacked Parametric Survival with Frailty Components and  Interval Censored Failure Times}
\author{\large{{Matthew W. Wheeler}$^1$, {Joost Westerhout}$^2 \dagger$} \\
 \large{ {Joe L. Baumert}$^3$ {Benjamin C. Remington}$^2 \dagger$} \\
\small{$^1${\textbf{Risk Evaluation Branch}}},\\ 
\small{National Institute for Occupational Safety and Health},  \\
\small{Cincinnati, OH} \\\small{\textbf{mwheeler@cdc.gov}} \\
\small{$^2$\textbf{TNO}}\\
\small{ Zeist, Netherlands} \\ 
\small{$^3$\textbf{University of Nebraska-Lincoln},} \\
\small{ Department of Food Science and Technology},\\
\small{ FARRP, Lincoln, NE 68588-6207}\\
\small{$\dagger${Equally contributing authors.}}
}

\begin{document}
\maketitle
\begin{abstract}
To better understand effects of exposure to food allergens, food challenge studies are designed to slowly increase the dose of an allergen delivered to allergic individuals until an objective reaction occurs. These dose-to-failure studies  are used to determine acceptable intake levels and are analyzed using parametric failure time models.   Though these models can provide estimates of the survival curve, their parametric form may misrepresent the survival function for doses of interest, and different models that describe the data similarly may produce different dose-to-failure estimates.  Motivated by predictive inference, we developed a Bayesian approach to combine survival estimates based upon posterior predictive stacking, where the weights are formed to maximize posterior predictive accuracy.  The approach allows for the inclusion of flexible models, and, in our case, allows us to include random effects to account for frailty components entering the model through study-to-study heterogeneity. The methodology is investigated in simulation, and is used to estimate allergic population eliciting doses for multiple food allergens. 

\end{abstract}

\section{Introduction}


Interval censored failure time data can be found in many areas of research including animal carcinogenicity experiments, demographical, epidemiological, financial, medical, sociological, machine reliability, and injury risk studies (Collett, 2015; Crowder et al., 1994; Klein et al., 2016; Ozturk et al., 2018; Sun, 2006; Yoganandan et al., 2016). Typical examples of interval censored data are found in medical or health studies with periodic follow-up protocols, with examples including time-to-tumor or time-to-infection studies (e.g. appearance of lung cancer tumors; breast cancer studies with 4-6 month follow-up times; timing of HIV infection in AIDS cohort studies with follow-up every 6 months) (Sun, 2006). Another example of interval censored failure time data can be found in the dosing steps of an oral food challenge studies to diagnose food allergy \cite{taylor2009}. However, the previously available methods for parametric modelling of interval censored failure data have been limited to single model approaches. In other fields of risk assessment, i.e. chemical risk assessment, single model selection and rejection has been acknowledged as a suboptimal approach and a more advanced, state-of-the-art method preferred for parametric modelling is model averaging (EFSA Scientific Committee, 2017; US EPA et al., 2019). Current model averaging methods available for chemical benchmark dose risk assessment are limited to a single dose or time point and corresponding model averaging techniques for interval censored data have not been previously available. In this study, we present a novel Bayesian methodology developed for interval censored failure data to combine parametric survival estimates with frailty components based upon posterior predictive stacking, where the weights are formed to maximize posterior predictive accuracy. The case study of oral food challenge data from food allergic patients is utilized to demonstrate the utility of the method.  

\begin{figure}
\centerline{\includegraphics[width=342pt,height=20pc]{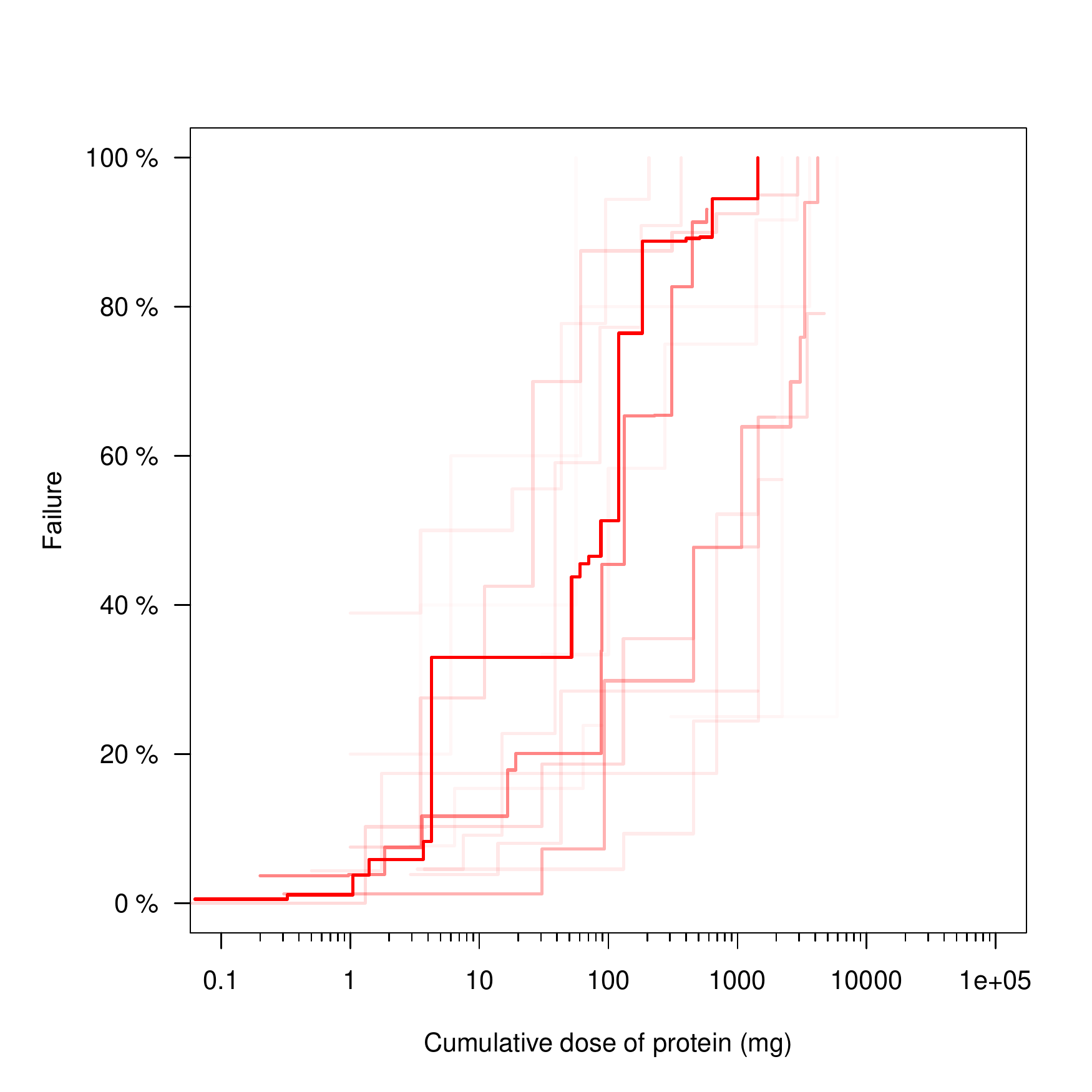}}
\caption{Individual Kaplan-Meier curves for 17 food challenge studies investigating Peanut protein 
			sensitized populations. The darker red curves represent studies with more observations.} \label{fig1}
\end{figure}

In studying food allergies, oral food challenge studies slowly increase the food protein dose until the individual presents allergic symptoms to a specific exposure amount (mg food protein).  Figure (\ref{fig1}) shows an example of such data for peanut allergies and represents 17 different clinical studies. Each study can range in size, from a small (10 to 30 participants) to a large (over 200 participants) dose-to-failure experiment with participants who present with objective, externally observable symptoms.  To more accurately estimate the survival curve,  separate studies are combined to estimate an exposed dose that represents a predefined, specific risk to the population. Risk assessors are typically interested in finding allergic population-based eliciting doses (EDs) where less than  $1\%$ or $5\%$ of the allergic population would experience an objective allergic reaction (ED01, ED05 respectively).   Such data are most appropriately analyzed using survival techniques \citep{crevel2007, taylor2009, taylor2014,dano2015} , but to accurately represent the population's dose-to-failure distribution, especially in the lower tails, use of a single survival distribution presents several challenges.  

Most critically, we are concerned with survival regression methods that provide an accurate representation of the dose-to-failure curve along with appropriate uncertainties being accurately represented. Traditionally, a non-parametric estimator, such as the Kaplan-Meier \citep{turnbull1976}, can be used for this task. However, food challenge data is not readily available for all foods, geographic locations, age groups or other sub-populations.  Thus, the data sets of interest can typically rely on less than $100$ observations where the true dose-at-failure is only known to be in an interval (i.e., the data are interval censored).  Non-parametric estimators, based on such limited data, may not be able to accurately capture dose-to-failure distribution in the tails of the failure distribution. 

Parametric failure time models are an alternative to non-parametric survival models if they fit the data well. In many cases, this is not the case, and these models may not be flexible enough, especially in the tails of the failure distribution, to describe a given data set.  As a consequence, estimation using any single model may result in bias and estimates whose parameters do not fully reflect the true uncertainty. With this in mind, a number of authors have considered flexible parametric survival models using  mixing distributions. Lambert \textit{et al.} \cite{lambert2004} used a mixture distribution consisting of a Gompertz hazard function and a single accelerated failure time (AFT) component.  Komarek and Lesaffare \cite{komarek2008} developed an AFT model with an over-specified number of mixture components. In both cases, the resulting time to failure distribution is  a convex sum of individual failure curves, which is more flexible than a single parametric model, but these approaches develop a new class of models to describe the data,  under the assumption the increased flexibility will accurately describe the observed data.  As a consequence, given appropriate data, the methodologies may suffer from the same issues as single failure time distributions.     Instead of creating a new survival model, we take the approach that it is possible to estimate the survival function as a weighted sum of the posterior distribution of existing parametric failure models.  As our approach is a general methodology it could incorporate any failure model, which includes those of Lambert \textit{et al.} and Komarek and Lesaffare. 

Unlike the above mixture approaches, we estimate failure distributions individually and focus on estimation where the failure distribution is representable as a convex sum of these estimates.   This is often accomplished using Bayesian model averaging \citep{raftery1997,hoeting1999} (BMA). However, when the true data generating mechanism is not in the suite of models considered BMA may not be appropriate. BMA is known to choose the model that is closest in Kullback-Leibler divergence to the observed data generating  mechanism \citep{yao2018}, which implies that the estimate will asymptotically be from one model.   In our problem,  there is no expectation that any of the parametric survival curves represent the true survival function, implying the estimate will be chosen by a single incorrect model that does not fully incorporate all of the  uncertainty.  

Instead of picking the best model that represents the data, we are interested in finding a set of weights that optimize posterior predictive inference.  Stacking \citep{wolpert1992,breiman1996,leblanc1996} produces estimates based upon multiple models, but focuses on predicting new observations based upon some scoring rule. Here point estimates and uncertainty in that estimate are constructed using a weighted average of estimates, where weights are formed by minimizing the squared error scoring rule in a leave-one-out cross validation. Originally used in frequentist settings, stacking has been given a Bayesian motivation by Clyde and Iverson \cite{clyde2013}.  They show that weights can be interpreted as maximizing the expected utility under several loss functions for the prediction of new data; Le and Clarke \cite{le2017} show that the stacking solution is the Bayes solution asymptotically, but, these results limit stacking to averaging point predictions.  Yao \textit{et al.} \cite{yao2018} extend these results making stacking applicable to estimating the posterior predictive distribution. We apply  Yao \textit{et al.} to stack posterior predictive distributions estimated from parametric failure time modeling.  This results in a flexible methodology to estimate the failure distribution using any fit parametric model, where no single model is expected to get weight $1$ as the sample size increases.  The methodology is relatively straightforward to implement and extend using a new parametric failure model in STAN \citep{carpenter2017}, and we provide an R package that implements the modeling for the data we consider.  

As any parametric model can be used with this methodology, we include  frailty components through random effects to account for study-to-study heterogeneity.  This approach has a long history of use in both semi-parametric \citep{klein1992} and parametric \citep{vaida2000,yamaguchi2002,komarek2007,crowther2014}  time-to-event data to represent frailty components.   In our data,  our effects enter the model as study specific random effects, where the frailties are thought to arise through:  differing protocols, differing participant recruitment, differing cumulative dosing designs, differing symptom interpretations by clinicians and nurses, and an array of possible regional genetic or environmental differences present in each study's population.    In addition to frailty components with standard right- and left- censoring, we include the possibility that the failure is only known in an interval to reflect the complicated nature of our data problem.    This is a flexible approach to analyzing failure time data that arise from a diverse set of circumstances.

The manuscript is organized by describing the general model framework, studying the method through a simulation, and showing the utility of the method by applying it to two food allergen dose-to-failure databases. Example databases for the manuscript include a large peanut allergy food challenge database \cite{taylor2014} and much smaller sesame allergy food challenge database \cite{dano2015} formed from multiple studies at different centers.

\section{Stacked Survival Regression}

To model these data, assume there are $J$ studies and each study
has $n_j$ observations,  where $j= 1, \ldots, J$. For subject  $i $ $, i = 1, \ldots, n_j, $ in study $j,$  we observe a failure in the interval  $\tau_{(ij)} =[t_{i1},t_{i2}] = [a_{jk},a_{j(k+1)}]$,  where the set $\{a_{jk}\}_{k=1}^{K_j+1},$ $a_{jk} < a_{j(k+1)},$ represents $K_{j+1}$ study specific intervals where the failure can be observed, cumulative dose intervals in our case.  For  $k=1$, $a_{j1}= 0$  and for $k=K_j+1$ one has $a_{j(K_j+1)}= \infty.$  For $J$ studies, we observe $T=\{\tau_{(11)},\ldots,\tau_{(n_j1)},\ldots,\tau_{(1J)},\ldots,\tau_{(n_JJ)}\}$ with $n= \sum_{j=1}^J n_j$ total observations and wish to accurately estimate the underlying generating process.

Let $F(t|\mu(x),\lambda)$ be a  cumulative distribution function (CDF), with $\mu(x)$ a function of $x$  defining the location and $\lambda$ a scale parameter. 
Given the failure model $F(t|\mu(x),\lambda)$ and noting $S(t \mid \mu(x),\lambda,x) = 1 -F(t \mid \mu(x),\lambda)$, the observed data likelihood for $T$  is
\begin{align}
   \ell(T | \mu(x),\lambda ) &= \prod_{j=1}^j \prod_{i=1}^{n_j} \bigg\{ F(t_{i2} \mid \mu(x_{ij}),\lambda) - F(t_{i1} \mid \mu(x_{ij}),\lambda) \bigg\}, \label{likelihood} 
\end{align}
where $x_{ij}$ is a vector of covariates.  

For a given $F(t\mid \mu(x),\lambda),$ (\ref{likelihood}) forms a key component for inference on the failure distribution, and $\mu(x)$ may include fixed and random effects.  In our problem, $x$ is a vector of indicator variables that indicate the observation's study and is related to $\mu(x)$ through the identity  or log link of $Bx,$ where $B = (b_0,\ldots,b_j)'$ and $b_0$ is a fixed intercept. That is, for study $j$ given CDF $F(t|\cdot),$ $\mu(x_j) = b_0 + b_j$  or $\mu(x_j) = \mbox{exp}(b_0 + b_j),$ which is used to define a random effect model over the study centers.   

When random effects are included in the model, inference on the population average survivor function may be of interest. This quantity  is estimated by marginalizing over $\mu.$, i.e.,  
\begin{align}
S(t\mid \lambda,x) = \int S(t \mid \mu(x),\lambda) d \hspace{0.1mm} \mu.
\label{marginal}
\end{align} 
For many survival functions, analytic evaluation of this integral is not possible.  In what follows, we estimate it using Markov Chain Monte Carlo (MCMC) methods. 

The above describes a likelihood model on a specific failure distribution that can be used in Bayesian or frequentist inference.  Given a CDF, there is  no expectation that any parametric failure distribution will be representative of the underlying process.  In what follows, we use Bayesian stacking to estimate the failure distribution from multiple parametric models
that can be analyzed individually using (\ref{likelihood}) as well as prior assumptions on  using Bayesian stacking. 

\subsection{Bayesian Stacked Survival}
For inference based upon an individual survival function,  we place a prior over $\mu(x)$  and $\lambda$ and estimate the posterior
distribution using MCMC. Inference on a single model may under represent the true uncertainty around the failure distribution.  Consequently, we perform multi-model inference
using Bayesian stacking. 

Assume there are $M$ individual failure distributions, i.e., $F_m(t|\mu_m(x),\lambda_m)$,  $m = 1,\ldots,M$,  estimated using MCMC.  We wish to find an optimal predictive distribution for $F(t|x)$ and $S(t|x)$ based upon these $M$ models.  To do this, we use \citet{yao2018} to estimate a set of optimal weights that places $F(t|x)$ and $S(t|x)$ in the convex hull of the $M$ posterior predictive densities.  That is we find $\hat{w} = (\hat{w}_1,\ldots,\hat{w}_M)$ and estimate the predictive density for new failure time $\widetilde{t}$ as
\begin{align}
\hat{p}(\widetilde{t}|T,x) &= \sum_{m=1}^{M} \hat{w}_m p_m( \widetilde{t}|T,x), \label{posteriorpred}
\end{align}
where $p_m(\widetilde{t}\mid T,x)$ is the posterior predictive distribution for model $m.$

As discussed in \citet{yao2018}, the weights are found by  
\begin{align*}
	\hat{w} &=  \underset{w\in \Omega }{argmax} \frac{1}{n} \sum_{j=1}^J\sum_{i=1}^{n_j} log \bigg[ \sum_{m=1}^{M} w_m p_m(\tau_{ij}|T_{-(ij)})\bigg],
\end{align*}
where $p_m(\tau_{ij}\mid T_{-(ij)})$  is the leave-one-out predictive distribution  computed using PSIS-LOO \citet{vehtari2017} and $\Omega$ is the space weights such that $\sum_{m=1}^{M} w_m = 1.$ By estimating the posterior predictive distribution  in (\ref{posteriorpred}) a large number of plausible failure time distributions can be investigated, and uncertainty in the true failure distribution can be appropriately expressed.  Further, this can be readily implemented in off the shelf software, which allows for straightforward implementation of the approach.  
 
\subsection{Estimation}

Individual model estimation proceeds using MCMC and the No-U-Turn Sampler (NUTS) \citep{hoffman2014} implemented in STAN \cite{carpenter2017}.  These posterior samples are used to estimate the stacked survival function. With (\ref{posteriorpred}) sampled using MCMC, there are several survival functions that  may be of interest.   If it is believed that an individual study may be representative of a susceptible sub-population, i.e., a representative of a group that may be particularly susceptible to failure, then the expected survival distribution for a given study can be computed as 
\begin{align*}
 E\bigg[S_j(\widetilde{t} \mid T, x_j) \mid T\bigg] &= \sum_{m=1}^{M} \int I(t > \widetilde{t})\hat{w}_m \hat{p}_m(t \mid T,x_j) \mbox{d} \mu_{(j)} \mbox{d}\lambda_{(j)} \mbox{d}t,
\end{align*}
where $\mu_{(j)}$ and $\lambda_{(j)}$ is shorthand to represent sampled parameter vectors for all M models given the data $T$,  that is these values are available from the sampling output.  This gives the point wise posterior predicted survival curve for an individual study with information shared between studies through the parameters. 

To estimate the population average survival function, the frailty components should be accounted for, which is accomplished by integrating out the random effects, i.e. 

\begin{align*}
 E\bigg[S(\widetilde{t} \mid T, x) \mid T \bigg] &=  \sum_{m=1}^{M} \int I(t > \widetilde{t}) \hat{w}_m p_m( t|T,x) \hspace{1mm} \mbox{d} \mu_m \hspace{1mm} \mbox{d} \lambda_m \hspace{1mm} \mbox{d} t,
\end{align*}
which is estimable over MCMC iterations.  

All estimates form the point-wise survival curve estimates. To compute the survival curve over the interval $(0,A]$, we estimate the point-wise curve over a fine grid of points and use a spline function to interpolate between points.  Specific time-to-failure probabilities are then computed from this curve.  For example, in our application the eliciting dose ($ED$) causing a group or population to have $y$ percent failure, i.e. $EDy$ is computed from the spline estimate.  Further, posterior predictive summaries of this value , e.g., distribution quantiles, can also be computed similarly as above. 

\section{Bayesian Specification}
\subsection{Failure Distributions}
The above specification is general and may include any density; however, for this analysis, we choose each $F_m(\cdot)$ from $M$ distributions that are either common in parametric survival regression, i.e., the Weibull distribution and Log-Gaussian (Log-Normal) and Log-Logistic, or whose mean $\mu_m$ is easily represented as a linear or exponential function of $x;$ additionally we include the generalized Pareto and Log-Laplace, also known as the log double-exponential, distribution, which have heavy tails, and are not typically used in the survival literature.  These choices give a broad range of survival functions that are applicable to many contexts, but the methodology is easily extended to other models if necessary.  Table (\ref{table:1} ) gives the five parametric distributions chosen in this specification. All of the distributions are determined by a $\mu(x)$ and a scale parameter $\lambda$.   

\begin{table}[htbp]
\centering
 \begin{tabular}{lcc}
Distribution & $F[t|\mu(x),\lambda]$  &$ \mu(x)$  \\ \hline\hline 
Weibull & $1 - \mbox{exp}\bigg[-\bigg( \frac{t}{\mu(x)} \bigg)^{\lambda} \bigg]$&  $\mbox{exp}(-b x) $\\ 
Generalized Pareto & $\frac{\lambda}{\mu(x)}\bigg( 1+ \frac{t}{\mu(x)}\bigg)^{-(\lambda+1)}$ & $\mbox{exp}(b x)$\\
Log-Gaussian &$\Phi\bigg[\frac{\mbox{log}(t)-\mu}{\lambda} \bigg]$  &  $b x$\\
Log-Logistic    & $\frac{1}{1 + [t/\mu(x)]^{-\lambda}} $&  $\mbox{exp}(-b x)$ \\
Log-Laplace    &  $\frac{1}{2}\bigg[ 1 +\mbox{sgn}\bigg(\mbox{log}\{t\} - \mu(x) \bigg)\bigg(1 - \mbox{exp}\bigg\{-\mid \mbox{log}[y]-\mu(x) \mid/\lambda \bigg\}\bigg)\bigg]$    &  $b x$ \\
\end{tabular}
\caption{Models and parameters used as failure time distributions in the analysis,.} \label{table:1}
\end{table}

\subsection{Prior Specifications}

We place diffuse priors over a range of plausible shapes for each distribution defined above.  For any given analysis, the observed failure times are typically on different scales.   For example, cancer recurrence times may be in the tens of years where as  cumulative dose-to-failure data may be in the thousands of micro-grams;  specifying a prior for a given failure distribution $F_m(\cdot)$ on a given scale may not translate to another scale. To create a general approach,  we scale all failure times to be in  the $[0,1]$ interval. Diffuse priors are subsequently defined relative to this interval.   

Table (\ref{table:Priors}) describes the priors for the each model. For each survival function,  $b_0$ defines mean of the random effects distribution placed over $b$ in Table (\ref{table:1}), where $b\sim N(b_0,z)$ and $z$ is a hyperparamter on the variance of the random effect specific to the distribution.  For a given model,  the dispersion parameter $\lambda$ is given a prior that is  diffuse over the interval $[0,1]$. For example in the Weibull model, the shape parameter  $\lambda$  is  log normal allowing a large range of shapes in this range, but for the Log-Gaussian this parameter is given a $\mbox{Gamma}(1,1)$ distribution. 

\begin{table}
\centering
\begin{tabular}{ccccc}
Model  & $b_0$  & $b$  & z  & $\lambda$ \\
\hline \hline
Weibull & $\mbox{Cauchy}(0,1)$ & $\mbox{N}(b_0,z) $ & $\mbox{IG}(1,1)$  & $\mbox{LN}(0, \sqrt{0.25}))$  \\
Generalized Pareto &$\mbox{Cauchy}(0,1) $  & $N(0,\sqrt{z})$  &$\mbox{IG}(1,1)$  & $\mbox{Gamma}(2,1)$   \\
Log-Gaussian &$\mbox{Cauchy}(0,1)$   & $\mbox{N}(b_0,\sqrt{z})$   & $\mbox{IG}(1,1)$  & $\mbox{Gamma}(1,1)$ \\
Log-Logistic &$\mbox{Cauchy}(0,1)$  & $\mbox{N}(b_0 ,\sqrt{z} )$ &  $\mbox{IG}(0.1,0.1)1_{[0 <  z  < 4]}$ & $\mbox{LN}(0,1)$ \\
Log-Laplace &$\mbox{Cauchy}(0,1)$ &  $\mbox{N}(b_0, z)$ & $\mbox{IG}(0.1,0.1)1_{[0<z<4]}$  &$\mbox{LN}(0,1)$ \\
\end{tabular}
\caption{Priors defined for each failure model. Distributions are specified such that $N(\mu,\sigma)$ represents the normal distribution with mean $\mu$ and standard deviation $\sigma$, $\mbox{IG}(a,b)$ is the inverse-gamma distribution with parameters $a$ and $b$,  and $LN(\mu,\sigma)$ is the log-normal distribution with log-mean $\mu$ and standard-deviation $\sigma.$ 
Additionally, $1_{[c< x <d]}$ is an indicator function truncating the random variable $x$ to be in the interval $[c,d].$} \label{table:Priors}
\end{table}

\section{Simulation}
We investigated the performance of this approach through simulation. For this study, we use the priors defined above, and, as we are concerned with inference in the tail of the dose-to-failure curve for food allergy related purposes, we look at estimating the doses associated with $0.01,0.05,$ and $0.10$ failure rates.  We compare the proposed approach to a Weibull model with a random effect term specified for each study.  

For the simulation, data are generated from two underlying true failure distributions, 
\begin{align*}
    F_1(d|\mu(x)_w,\mu(x)_{IG},) &= 0.7\hspace{1mm} \mbox{Weibull}(\mu(x)_w,10) + 0.3\hspace{1mm} \mbox{IG}(\mu(x)_{IG}, 0.25) \\
    F_2(d|\mu(x)_{IG},\mu(x)_{ST}) &= 0.5 \hspace{1mm} \mbox{IG}(\mu(x)_{IG}, 0.25) + 0.5 \hspace{1mm}\mbox{log-SkewT}(\mu(x)_{ST},1,3),\\
\end{align*}
which represent mixture distribution from two distinct failure models dependent on study $x$. Here,  $\mbox{IG}$ is the CDF of the inverse-Gaussian distribution having expectation $\mu(x)_{IG}$ and scale $\mbox{0.25},$ and $\mbox{log-SkewT}$ is the CDF of log of the skew-T distribution, that is if $D \sim \mbox{log-SkewT}(\mu(x)_{ST}, 1,3)$ then $\mbox{log}(D) \sim \mbox{SkewT}(\mu(x)_{ST}, 1,3)$ where $\mu(x)$ is the center parameter, $1$ is the dispersion parameter, and 3 is the skew parameter.   By defining the failure distribution to be a mixture distribution, the study ensures the true failure distribution is not in the suite of distributions fit to the data; however, $F_1(d|\cdot)$ is close to the Weibull model that is fit to the data as a comparison.  This allows a comparison of the robustness of inference  when the model is slightly miss-specified.   Further, $F_2(d|\cdot)$ is a case where the model is completely miss-specified, which allows a comparison of the methodologies in cases expected to arise in practice. 

To accommodate study-to-study variability,  multiple study centers are included in each simulation. For each iteration and distribution/study, $\mu(x)$ is drawn from a  log-Gaussian distribution with log-standard deviation of  $0.25$ and log-mean of  $1,7$ and $-0.8$ for the Weibull, inverse-Gaussian, and log-Skew T distributions respectively.  This process was done 200 times for n = $5,15,30$  study centers.  For each study center, a random number of subjects were observed, where the total number of subjects was drawn from a Poisson distribution with mean $10$.  Additionally, the interval censoring locations were considered random for each study.  Here 10 dosing regimes were drawn from a log-Gaussian distribution with log mean of 1 and log-standard deviation of 1.4. To ensure these doses were not close to zero, i.e. there was appreciable positive probability of a left censoring event, 0.75 was added to each tested dose. The dose-to-failure was estimated for probabilities $1, 5,$ and $10\%$, and the mean squared integrated error for the proposed approach and the Weibull approach are computed.

Table (\ref{tbl:Table3}) gives the ratio of the mean squared error (MSE) between the two approaches. Values less than one indicate the proposed approach has less mean squared error, while asterisks indicate that ratio is significantly different at the $\alpha = 0.05$ level. In the case of the inverse-Gaussian/log-Skew T case, which is the case the model is completely mis-specified, the proposed approach has significantly less MSE than the Weibull approach.  Further, the improvement increases as the number of study centers increases, which indicates that the difference in the MSE improves as more study centers are added. In the Weibull/inverse-Gaussian case,
a more nuanced result is seen. Here the MSE is approximately the same between the two approaches, when $n=5$ and $n=15,$ but does favor fitting the single Weibull model when $n=30.$  for smaller sample  with the Weibull random effects model appearing to have slightly less MSE although this value is never significantly different at the $0.05$ level across the simulation conditions.

\begin{table}[htbp]
  \centering
     \begin{tabular}{ccccccc}
                 & \multicolumn{3}{c}{Inverse-Gaussian/Skew-T} & \multicolumn{3}{c}{Weibull/Inverse-Gaussian} \\
                 & \multicolumn{3}{c}{Number of study centers} & \multicolumn{3}{c}{Number of study centers} \\
   Failure Probability       & \textbf{5} & \textbf{15} & \textbf{30} & \textbf{5} & \textbf{15} & \textbf{30} \\\hline \hline
           $1\%$  & 0.53* & 0.43* & 0.29*   & 1.05  & 1.02  & 1.33 \\
           $5\%$  & 0.59* & 0.47* & 0.38*   & 0.99  & 0.98  & 1.07 \\
           $10\%$ & 0.76* & 0.68* & 0.62*   & 0.98  & 1.02  & 1.30 \\
    \end{tabular}%
 	 \caption{Ratio of mean squared error of the proposed approach to the Weibull random effect model when estimating the dose that causes  failure at $1, 5$ and $10\%$.  Asterisks indicate the ratio is different from $1$ at the $0.05$ level. }
 \label{tbl:Table3}%
\end{table}%

\section{Food Allergy Data }

We apply the proposed approach to study two cumulative dose-to-failure failure distributions
for populations exposed to sesame and peanut allergies.  These databases and corresponding frequentist analyses using accelerated failure time models are described in \citet{taylor2014} and \citet{dano2015}.   For each individual study in a given database, known or suspected allergic participants were recruited and exposed to increasing doses of the allergenic food protein. For a given dose, if no objective symptoms were observed after a set time period, the dose was increased to the next dosing step until the first objective symptom(s) of an allergic reaction was observed or the maximum predetermined experimental dose was reached.  When objective symptoms were presented, the study was stopped, and the participant was marked as having an objective allergic reaction between the previous cumulative dose with no objective symptoms (NOAEL, No Observed Adverse Effect Level) and the current cumulative dose with observed objective symptoms (LOAEL, Lowest Observed Adverse Effect Level). Each individual study in a given database recruited participants from diverse geographic populations and followed its own internal protocol determining the nature and severity of symptoms required to terminate the experiment.  As a result, individual studies exhibit marked heterogeneity. We estimate the entire failure distribution and utilize the entire distribution in certain quantitative risk assessment scenarios. However, for the purposes of derivation of eliciting doses which possibly could be used later as  Reference Doses such as those proposed the the Allergen Bureau of Australia and New Zealand's Voluntary Incidental Trace Allergen Labelling (VITAL) program \cite{taylor2014}, we are interested in the cumulative protein dose that would result in an objective allergic reaction in less than $1\%$ or $5\%$ of the allergic population (i.e., ED01 or ED05) which is important for risk assessment purposes.  For our purposes, the two databases analyzed represent a small database (three studies with less than 40 participants for sesame) and a large database (17 studies with more than 700 participants for peanut) to show the utility of the proposed approach. Analyses for these databases use the same prior distributions as those in the simulation study. 

\subsection{Peanut dose-to-failure}
 \citet{taylor2014} reported 750 peanut-allergic individuals with 30 left-censored and 132 right-censored datapoints. Using accelerated failure time models, cumulative ED01 values were reported as 0.13 mg peanut protein for the Log-Logistic distribution and 0.28 mg peanut protein for the Log-Gaussian distribution. Accelerated failure time ED05s were 1.4 and 1.5 mg peanut protein, respectively, with ED10s of 4.1 and 3.8 mg peanut protein. Based upon the available data and expert judgement, there was no biological reason to distinguish between the results of either distribution and the VITAL Scientific Expert Panel recommended a Reference Dose of 0.2 mg peanut protein to guide application of precautionary allergen labeling and achieve the desired protection level for $99\%$ of the the allergic population \cite{taylor2014}. However, this Reference Dose has come into question as the process did not rely on a single dose distribution. Reanalysis of the \cite{taylor2014} peanut database with the proposed approach in this manuscript results in a stacked survival estimate  ED01 of 0.2 mg peanut protein, an ED05 of 3.3 mg peanut protein and an ED10 of 10.9 mg peanut protein. Unsurprisingly, the stacked survival estimate  values are quite close to the previous accelerated failure time results, and the stacked survival estimate  ED01 of 0.2 mg peanut protein is the same as the previously suggested \cite{taylor2014} VITAL reference dose of 0.2 mg peanut protein. The  Weibull, Generalized Pareto and Log-Laplace distributions were the 3 models giving the most weight to the ED estimation for the peanut reanalysis. The Generalized Pareto and Log-Laplace distributions were not previously available for \cite{taylor2014} or \cite{dano2015}, demonstrating the added value of the new approach and reinforcing the conservative nature of the previous estimates.

\begin{figure}
    \centerline{\includegraphics[width=342pt,height=20pc]{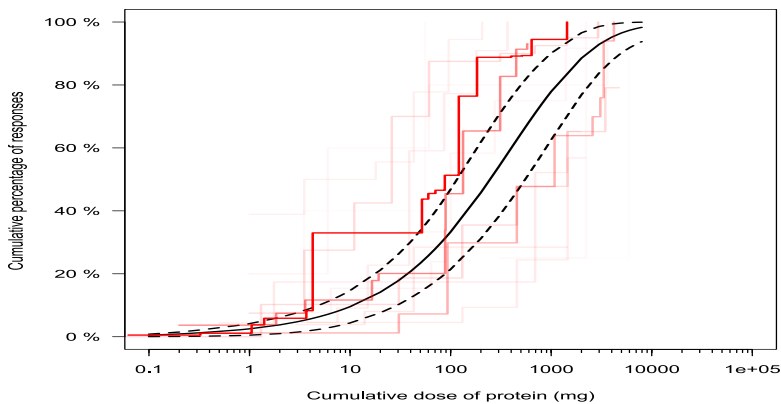}}
    \caption{Stacked posterior predicted failure distribution from 17 food challenge studies estimating peanut protein allergies. Center black line is the stacked central estimate, and the dashed black line represents the estimated $90\%$ posterior predicted failure times for the population. Red lines represent individual Kaplan-Meier curves for the studies. The darker red curves represent studies with more observations.} \label{fig3}
\end{figure}

\subsection{Sesame dose-to-failure}

\cite{dano2015} reported 35 sesame-allergic individual datapoints. Using accelerated failure time models, the lower $95\%$ confidence interval estimates for the cumulative ED05 values were reported as 0.4 mg sesame protein for the Log-Logistic distribution, 0.6 mg sesame protein for the Log-Gaussian distribution, and 0.1 mg sesame protein for the Weibull distribution. Accelerated failure time ED05s were 2.1, 2.4, and 1.0 mg sesame protein, respectively. Based on a similar but slightly smaller database, the VITAL Scientific Expert Panel utilized the lower $95\%$ confidence interval of the ED05 to recommend a Reference Dose of 0.2 mg sesame protein \cite{taylor2014}. Reanalysis of the \cite{dano2015} sesame database with the proposed approach in this manuscript results in a stacked survival estimate  ED05 of 5.1 mg sesame protein, with a lower $95\%$ confidence interval of 0.6 mg sesame protein. The stacked survival estimate  values are quite close to the previous accelerated failure time results, and the stacked survival estimate of the lower $95\%$ confidence interval estimate for the ED05 of 0.6 mg sesame protein is slightly higher that the previously suggested reference dose of 0.2 mg sesame protein, and it is in 0.1 to 0.6 mg sesame protein range from the 3 accelerated failure time model estimates. The  Log-Laplace distribution, which was not previously available for \cite{taylor2014} or \cite{dano2015}, was model giving the most weight to the ED estimation for the sesame reanalysis, again demonstrating the added value of the new approach. 

\begin{figure}
    \centerline{\includegraphics[width=342pt,height=20pc]{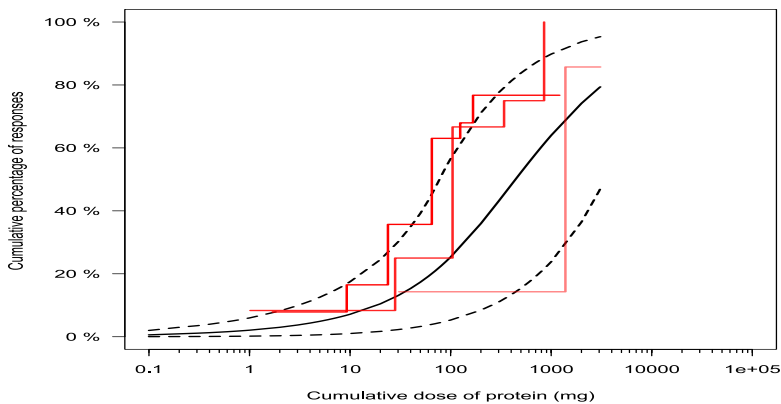}}
    \caption{Stacked posterior predicted failure distribution from 3 food challenge studies estimating sesame protein allergies. Center black line is the stacked central estimate, and the dashed black line represents the estimated $90\%$ posterior predicted failure times for the population. Red lines represent individual Kaplan-Meier curves for the studies, and the grey line is the estimated Kaplan-Meier curve for the entire database. The darker red curves represent studies with more observations.} \label{fig2}
\end{figure}

\section{Discussion}
We have presented a general framework for survival regression based upon stacking, and have applied it to several dose-to-failure data-sets where it is of interest to determine levels of exposure that do not typically elicit a response from a majority of the sensitized population.   Although we have  assumed random effects for our model to account for study-to-study variability, other modifications of this approach can be envisioned.  For example, if necessary, it is possible given the framework to include fixed effects.  In such a case, survival functions given these covariates can be  constructed from the posterior predictive distribution. 

The major benefit of this approach is that it can incorporate any parametric failure distribution.  Extensions to the model are straight forward, and individual models can be added to the code provided in the appendix.  Flexible survival models, such as those described by \citet{lambert2004} can be included providing an additional layer of  flexibility. Finally, though we have shown the applicability of the methodology with regard to clinical data from the field of food allergy, it is anticipated that the method has utility outside of this domain. Additional possible domains of application could include medical related time-to-tumor research, carcinogenicity experiments in animals, large epidemiological studies, machine maintenance studies, estimation of dietary exposure to contaminants or other fields where interval censored data is present. For example, in the machine reliability literature, where accurate estimates of the failure times in the tail are important, this methodology may be a useful addition.  

\section*{Acknowledgements}
The authors would like to thank Randall Smith and Drs Dustin Long and Michael Pennell for comments on an earlier version of the manuscript. The findings and conclusions in this report are those of the authors and do not necessarily represent the official position of the National Institute for Occupational Safety and Health, Centers for Disease Control and Prevention. 

\bibliography{bibliography}

\end{document}